# Tripartite Perspective on the Copyright-Sharing Economy in China

Jyh-An Lee[*]

Faculty of Law, The Chinese University of Hong Kong

## Abstract

Internet and digital technologies have facilitated copyright sharing in an unprecedented way, creating significant tensions between the free flow of information and the exclusive nature of intellectual property. Copyright owners, users, and online platforms are the three major players in the copyright system. These stakeholders and their relations form the main structure of the copyright-sharing economy. Using China as an example, this paper provides a tripartite perspective on the copyright ecology based on three categories of sharing, namely unauthorized sharing, altruistic sharing, and freemium sharing. The line between copyright owners, users, and platforms has been blurred by rapidly changing technologies and market forces. By examining the strategies and practices of these parties, this paper illustrates the opportunities and challenges for China's copyright industry and digital economy. The paper concludes that under the shadow of the law, a sustainable copyright-sharing model must carefully align the interests of businesses and individual users.

---
[*] Correspondence to: Faculty of Law, The Chinese University of Hong Kong, Shatin, New Territories, Hong Kong.
*E-mail address*: jalee@cuhk.edu.hk

**\*434**

## 1. Introduction

By facilitating an unprecedented scale and speed of the flow of information, the Internet and digital technologies have enabled a new sharing economy, turning both tangible and intangible assets into sharable resources and unlocking their value for people who want them.[1] In other words, such a sharing economy has been enabled by the fact that content can be digitalized and distributed at near-zero marginal cost.[2] However, because intangible assets shared online are often copyrighted and the sharing economy has swept across various copyright domains, such as music, film, computer software, and 3D printing,[3] this new economy has created unparalleled challenges for the protection and enforcement of copyright.[4] **\*435** It is widely recognised that the primary policy goal underlying a copyright regime is to maintain a delicate balance between authors' incentive to create and users' interest in accessing the copyright works.[5] As yet, it is unclear whether the sharing economy facilitated by digital networks will help maintain this balance or disrupt it in a fundamental and unforeseen way.

Although the copyright-sharing economy is sometimes viewed as similar to the sharing economy associated with physical assets,[6] such as car sharing and home sharing, the nature of the copyrighted information distinguishes the copyright-sharing economy from other forms of the sharing economy. In contrast to tangible assets, the information protected by intellectual property (IP) laws has strong public-goods characteristics, namely non-rivalrousness and non-excludability.[7] Information is non-rivalrous because the same piece of information can be consumed by multiple parties at the same time.[8] One's consumption of certain information would not diminish the amount of the same information available to others.[9] In contrast to physical property, information can never be worn out, crowded, or impaired as a result of being shared.[10] Information is also non-excludable, because it is often quite challenging to prevent its dissemination.[11] The public-goods nature of information has made it a reasonably suitable object for sharing.

---

[1] Yochai Benkler, 'Sharing Nicely: On Shareable Goods and the Emergence of Sharing as a Modality of Economic Production' (2004) 114 Yale LJ 273, 275-76; Giancarlo F. Frosio, 'Resisting the Resistance: Resisting Copyright and Promoting Alternatives' (2017) 23 Rich J L & Tech 4, 5; Mark A. Lemley, 'IP in a World without Scarcity' (2015) 90 NYU L Rev 460, 494; Orly Lobel, 'The Law of the Platform' (2016) 101 Minn L Rev 87, 88-89; Jessica Litman, 'Sharing and Stealing' (2004) 27 Hastings Comm & Ent LJ 1, 2.
[2] Eric Priest, 'Meet the New Media, Same as the Old Media: Real Lessons from China's Digital Copyright Industries' (2016) 23 Geo Mason L Rev 1079, 1080.
[3] Matthew David, *Sharing: Crime Against Capitalism* (Polity 2017) 5-6.
[4] Ben Depoorter, 'Technology and Uncertainty: The Shaping Effect on Copyright Law' (2009) 157 U Pa L Rev 1831, 1848; Ryan Wichtowski, 'Increasing Copyright Protection for Social Media Users By Expanding Social Media Platforms' Rights' (2017) 15 Duke L & Tech Rev 253, 253-54.
[5] Charlotte Waelde and Hector MacQueen 'From Entertainment to Education: The Scope of Copyright?' (2004) 3 IPQ 259, 268-69.
[6] Daniel K. McDonald, 'Is the Sharing Economy Taxing to the Traditional?' (2017) 16 Fla St U Bus Rev. 73, 74-75; Lobel (n 1) 126-42; Inara Scott and Elizabeth Brown, 'Redefining and Regulating the New Sharing Economy' (2017) 19 U Pa J Bus L 553, 563.

However, IP represents a legal monopoly conferred by the law, which has made IP-protected information exclusive.[12] Parties other than copyright owners are generally not allowed to legally share copyrighted works without licenses. The exclusivity feature of IP law has frequently created impediments for the sharing economy. Therefore, some have argued that the exclusive copyright is inconsistent with the society's interest in creating new copyright works through sharing.[13] Paradoxically, although the exclusive nature of IP seems to contradict the principle of sharing, sharing has served as an underlying idea of copyright philosophy. Every author's work must build upon the past and borrow ideas developed by others.[14] The copyright system cannot facilitate continuous creativity without providing access to the work of previous creators. The delicate relations between copyright and the sharing economy have therefore given rise to quite a few unsolved issues in law, innovation, and business strategy.

China is an interesting and valuable context in which to observe copyright challenges brought about by the sharing economy because of its longstanding piracy

---

[7] Michele Boldrin and David K. Levine, *Against Intellectual Monopoly* (CUP 2010) 156; William W. Fisher III, *Promises to Keep: Technology, Law, and the Future of Entertainment* (Stanford University Press 2004) 199-200; Eric E. Johnson, 'The Economics and Sociality of Sharing Intellectual Property Rights' (2014) 94 BU L Rev 1935, 1940.

[8] Johnson (n 7) 1940-41; Lemley (n 1) 466-67.

[9] Yochai Benkler, *The Wealth of Networks: How Social Production Transforms Markets and Freedom* (Yale University Press 2006) 36; Boldrin and Levine (n 7) 156;David (n 3) 9; William M. Landes and Richard A. Posner, *The Economic Structure of Intellectual Property Law* (Harvard University Press 2003) 19-20.

[10] Ibid. 13-14.

[11] Fisher (n7) 200; Johnson (n 7) 1940.

[12] Benkler (n 9) 36-37; Lemley (n 1) 467.

[13] Raymond Shih Ray Ku, 'The Creative Destruction of Copyright: Napster and the New Economics of Digital Technology' (2002) 69 U Chi L Rev 263, 268.

[14] Joelle Farchy, 'Are Free Licences Suitable for Cultural Works?' (2009) 31 EIPR 255, 260

problem[15] and its extremely fast-growing electronic-commerce industry.[16] China's IP policy is now at the crossroads of imitation and innovation.[17] The country has the largest population of Internet users in the world. China now have more than 800 million Internet users.[18] Information sharing has been viewed as an important cultural element in the history of Chinese civilisation.[19] The Chinese sharing philosophy can be traced back to the ancient philosopher Mencius, who said that 'it is more joyful to share the joy than to keep it to oneself (独乐乐不如众乐乐)'.[20] Persuading King Qi Xuan to share his music with more people, Mencius eloquently argued that music is more enjoyable when it is heard with company.[21] To Mencius, sharing was an important way for the king to obtain and solidify support from *436 Chinese society.[22] After 2300 years, music sharing in China remains a hot issue, but it is now shaped by the new context of IP protection, innovation, entertainment, and economic development.

This paper describes and analyses the copyright-sharing ecology of China from the perspectives of three major players, namely copyright holders, users, and online platforms. Although there are other important players, such as search engines, peer-to-peer (p2p) file-sharing software providers, and copyright collective societies, these three stakeholders and their relations form the main structure of the copyright-sharing economy. The incentives for and strategies adopted by each stakeholder have a tremendous impact on the efficiency and success of the copyright-sharing system. Moreover, the dynamics of law, market activities, and technology associated with these stakeholders deserve more academic attention.

---

[15] William P. Alford, *To Steal A Book Is An Elegance offense: Intellectual property Law in Chinese Civilization* (Stanford University Press 1995) 86-87; Lucy Montgomery and Eric Priest, 'Copyright in China's Digital Cultural Industries' in Michael Keane (ed), Handbook of Cultural and Creative Industries in China (EE 2016) 342-44; Andrew C. Mertha, *The Politics of Piracy: Intellectual Property in Contemporary China* (Cornell University Press 2005) 3-5, 46-47, 61-62, 219; Eric Priest, 'Copyright Extremophiles: Do Creative Industries Thrive or Just Survive in China's High-Piracy Environment?' (2014) 27 Harv J L & Tech 467, 473-81; Diming Tang and Robert Lyons, 'An Ecosystem Lens: Putting China's Digital Music Industry into Focus' (2016) 1(4) Global Media and China 350, 352, 357-58, 360, 364; Susan Tiefenbrun, 'A Hermeneutic Methodology and How Pirates Read and Misread the Berne Convention' (1999) 17 Wis Int'l LJ 1, 3.

[16] Efraim Turban, Jon Outland, David King, Jae Kyu Lee, Ting-Peng Liang, Deborrah C Turban, *Electronic Commerce 2018: A Managerial and Social Networks Perspective* (Springer 2018) v; Dong Han, '"Use" Is an Anagram of "Sue": Cultural Control, Resistance, and the Role of Copyright in Chinese Cyberspace' (2011) 7(2) Global Media & Communications 97, 97-98.

[17] Yahong Li, Imitation to Innovation in China: The Role of Patents in Biotechnology and Pharmaceutical Industries (EE 2010).

[18] Niall McCarthy, 'China Now Boasts More Than 800 Million Internet Users And 98% Of Them Are Mobile' Forbes 2018 <https://www.forbes.com/sites/niallmccarthy/2018/08/23/china-now-boasts-more-than-800-million-internet-users-and-98-of-them-are-mobile-infographic/#63574ec27092> accessed 21 January 2019; 'China Focus: China Has 802 Million Internet Users' Xinhua News 2018 <http://www.xinhuanet.com/english/2018-08/21/c_137405424.htm> accessed 21 January 2019. 'China: Number of Internet Users December 2017' Statista <https://www.statista.com/statistics/265140/number-of-internet-users-in-china/> accessed 21 January 2019.

[19] Yahong Li and Graham Greenleaf, 'China's Copyright Public Domain: A Comparison with Australia' (2017) 27 AIPJ 147, 150.

[20] James A. Ryan, 'Moral Philosophy and Moral Psychology in Mencius' (1998) 8(1) Asian Philosophy 47, 53.

[21] Marthe Chandler, 'Meno and Mencius: Two Philosophical Dramas' (2003) 53(3) Philosophy East and West 367, 371.

[22] Ibid.; Michael Nylan, 'On the Politics of Pleasure' (2001) 14(1) Asia Major 73, 77-78.

Based on the case of China, this paper aims to contribute to the legal scholarship on a number of issues. First, as explained previously, IP sharing is different from the sharing economy based on offline competition, such as car sharing and p2p accommodation. The public-goods character of information distinguishes IP sharing from other types of the sharing economy. Second, this paper explains the consistency and inconsistency between the nature of IP and that of the sharing economy. The sharing economy seems to operate under an open ideology, which stands in opposition to the proprietary nature of IP. Nevertheless, with respect to software development and the music industry in China, we found that the sharing economy is either built upon the IP regime or provides an alternative to IP for copyright owners to extract value from their works. Third, this study conceptualizes current copyright sharing economy in China into three categories: unauthorized sharing, altruistic sharing, and premium sharing. Based on these three categories, this paper further explores the rationale behind the copyright practice of right holders, users, and platforms. Fourth, I analyse a uniquely Chinese phenomenon in which online platforms gradually become the major copyright owners in the video-streaming market. The changing business model has created more incentives for major Chinese online platforms to protect copyright. The content shared on those platforms has consequently become more limited and such sharing is designed as part of the freemium model. Last but not least, this paper investigates the role of government platforms in the copyright-sharing economy and its relevant policy implications. A careful observation of the government platforms' copyright policies may shed light on whether certain policy goals can be achieved by those platforms and their licensing strategies.

## 2. Conceptualizing copyright sharing economy

Copyright work may be shared by different stakeholders in different ways. Current copyright scholarship has not yet developed consensus regarding the definition of copyright sharing. Therefore, copyright sharing may refer to various types of content sharing in dissimilar contexts. In order to distinguish different copyright sharing models from one another, this section conceptualizes copyright sharing economy into three main categories. Sometimes those models mix with each other or transform into another. A correct understanding of the economic rationale underlying each sharing model is essential to capture the comprehensive picture of copyright sharing economy. The categorization will also be used in analysing copyright ecology in other sections of this paper.

### 2.1. Type I: Unauthorized sharing

Copyright sharing sometimes represents unlicensed sharing which constitutes copyright infringement.[23] Users might share others' copyright work via p2p file-sharing software or online platforms without copyright owners' permission.[24] Therefore, copyright owners are normally hostile against this type of sharing and have occasionally indicated that sharing is equivalent to stealing.[25] Negatively associating copyright sharing with theft and stealing has been frequently seen in the music and film industries' public relation campaigns.[26] Nonetheless, in order to balance copyright owners' private interests and the public's interests in accessing copyright works, copyright law provides exceptions and limitations under specific circumstances where unauthorized use or share of copyright works is not deemed infringement.[27]

## 2.2. Type II: Altruistic sharing

Certain copyright owners are willing to share their works for various altruistic reasons. Some creators build connections with online communities for further collaboration by sharing their works;[28] others may simply find it rewarding to share their works.[29] In the offline world, some communities have established strong norms by sharing creative works.[30] These sharing activities are characteristic of the gift economy, which is distinguished from the commodity economy, in which labour is supplied as a function of wages.[31] Contributors to free or **\*437** open-source software and Wikipedia are all examples of copyright owners sharing their works in the online gift economy,[32] where 'access to information is regulated not by price but by a complex set of social relations'.[33]

Nevertheless, copyright owners' incentives to share their works are occasionally quite complicated. Some right holders are willing to share their works not only for altruistic reasons but also to build their own reputation. For example, software developers may contribute their code to the free or open source software (F/OSS) projects for both the benefits of the community and the advancement of their own careers.[34] By solving complicated problems, programmers signal their abilities to others, including potential employees.[35] While altruistic sharing and freemium sharing may coexist in a hybrid model of sharing, the more monetary incentives underlying a sharing behaviour, the closer it will develop toward a freemium model introduced below.

---

[23] Enrico Bonadio, 'File Sharing, Copyright and Freedom of Speech' (2011) 33(10) EIPR 619, 621-22; Haflidi Kristjan Larusson, 'Uncertainty in the Scope of Copyright: The Case of Illegal File-sharing in the UK' (2009) 31(3) EIPR 124, 124, 130.

[24] Pheh Hoon Lim and Louise Longdin, 'P2P Online File Sharing: Transnational Convergence and Divergence in Balancing Stakeholder Interests' (2011) 33(11) EIPR 690.

[25] Peter J. Karol, 'Hey, He Stole My Copyright! Putting Theft on Trial in the Tenenbaum Copyright Case' (2013) 47 New Eng L Rev 887, 888, 890, 895, 898, 899; Daniel Lieberman, 'A Homerun for Three Strikes Law: Graduated Response and Its Bid to Save Copyright' (2012) 59 J Copyright Soc'y USA 223, 230; Litman (n 1) 23-24, 30.

[26] Ben Depoorter, Alain Van Hiel , and Sven Vanneste, 'Copyright Blacklash' (2011) 84 S Cal L Rev 1251, 1289-90; Karol (n 25) 896.

[27] Robert P. Merges and Seagull Haiyan Song, Transnational Intellectual Property Law: Text and Cases (EE 2018) 417-56.

[28] Lawrence Lessig, Remix: Making Art and Commerce Thrive in the Hybrid Economy (The Penguin Press 2008) 148.

[29] Johnson (n 7) 1958-59.

[30] Meredith G Lawrence, 'Edible Plagiarism: Reconsidering Recipe Copyright in the Digital Age' (2011) 14 Vand J Ent & Tech L 187, 205-12.

[31] Duran Bell, 'Modes of Exchange: Gift and Commodity' (1991) 20 J Socio-Econ 155, 163; Litman (n 1) 8.

[32] Jyh-An Lee, 'Organizing the Unorganized: The Role of Nonprofit Organizations in the Commons Communities' (2010) 50 Jurimetrics 275, 300-301.

[33] Lessig (n 28) 145.

## 2.3. Type III: Freemium sharing

The freemium model is commonly adopted for the marketing of various goods or services. Producers are willing to share their products for free because such sharing may bring about the potential massive-scale consumption of their products for a fee.[36] Sometimes users are provided with the basic features at no cost but will be charged for richer functionality.[37] In other words, freely shared products may generate users' interests in further purchasing the same products if their experience with the free product is satisfying. Freemium sharing model is widely defined in this paper to include individual creators who are willing to share their works to gain reputation or attention in the markets of both employment and copyright works.

Freemium sharing model has been increasingly important for the promotion of copyright products. This is first because of the fast and low-cost distribution of digitalized content as the distribution speed is usually the key to the success of any freemium model.[38] Producers' efficient provision of free samples and following-up paid products has facilitated consumers' continuous consumption.[39] Secondly, in order to profit from the freemium model, the producers need to reach a certain scale of consumption of which the revenue from paid products exceeds the cost of free ones.[40] Internet and digital technologies are also one of the main driving forces that help copyright holders reach the unprecedented scale of consumers. Third, copyright works, such as novels, music, and films, are typical experience goods.[41] Consumers do not know their value until they experience the products.[42] The parts that copyright owners share are the advertisement of the parts they sell.[43] Therefore, the freemium strategy can also be used to promote the reputation for new comers in the market.

---

[34] Josh Lerner and Jean Tirole, 'Some Simple Economics of Open Source' (2002) 50 J Industry Econ 197, 214.
[35] Ibid.
[36] Eric Benjamin Seufert, Freemium Economics: Leveraging Analytics and User Segmentation to Drive Revenue (Elsevier 2014) 1.
[37] Vineet Kumar, 'Making "Freemiun" Work' (2014) 92(5) Harvard Business Review 27, 27.
[38] Seufert (n 36) 1.
[39] Carl Shapiro and Hal R. Varian, Information Rules: A Strategic Guide to the Network Economy (Harvard Business School Press 1999) 86.
[40] Seufert (n 39) 3-4.
[41] Shapiro and Varian (n 39) 5, 85.
[42] Ibid. 85.

Among the three types of copyright sharing introduced above, only Type I (unauthorized sharing) is not initiated by copyright holders. Type II (altruistic sharing) and Type III (freemium sharing) may exist independently or co-exist over the same copyright work. There are certainly some other copyright-sharing activities that do not belong to any of these three categories or deviate from these three categories. For example, one may share his self-produced video via YouTube for altruistic or other non-monetary reasons in the beginning, but later becomes a successful YouTuber by earning millions of advertising revenue from such sharing.[44] Nonetheless, for the analysis of the Chinese copyright-sharing economy, this article only focuses on sharing activities of Types I to III.

## 3. Copyright owners

In contrast to the sharing economy with preexisting offline counterparts, such as Airbnb and Uber, owners of IP in the sharing economy have faced challenges in recouping their investments in property. First, the information protected by copyright is naturally nonappropriable.[45] Once the information is shared, it is quite difficult for its producer to appropriate its value through sale.[46] Consumers tend to free ride on the information at little cost.[47] Moreover, the Internet has dramatically reduced the cost of reproduction and dissemination of information. On one hand, copyright owners now have more channels through which to share their works;[48] on the other, it is increasingly difficult for them to control their work as they did in the past.[49] ***438**

Like those in other jurisdictions,[50] most established professional copyright owners in China have endeavoured to protect and enforce their copyright in the digital environment. For example, based on the famous director Chen Keigo's movie *The Promise*, Hu Ge created a 20-minute video titled *The Bloody Case That Started from a*

---

[43] Ibid. 86.
[44] Ian C. Butler, 'Note: The Ethical and Legal Implications of Ad-Blocking Software' (2016) 49 Conn. L. Rev. 689, 700; Hank Fisher, 'Comment: Danger in the DMCA Safe Harbors: The Need to Narrow What Constitutes Red Flag Knowledge' (2015) 49 U Rich L Rev 643, 654-55; James Puddington, 'Note: Fair Play: Economic Justifications for Applying Fair Use to the Online Streaming of Video Games' (2015) 21 BU J Sci & Tech L 413, 413-14.
[45] Marshall A. Leaffer, *Understanding Copyright Law* (LexisNexis 5th edn, 2010) 23.
[46] Ibid.
[47] Ibid.
[48] David (n 3) 3; Jessica Litman, *Digital Copyright* (Prometheus Books 2001) 108; Jessica Gutierrez Alm, '"Sharing" Copyrights: The Copyright Implications of User Content in Social Media'(2013) 35 Hamline J Pub L & Pol'y 105, 105; Eugenia Georgiades, 'The Limitation of Copyright: Sharing Personal Images on Social Networks' (2018) 40 EIPR 30, 30.
[49] David (n 3) 3; Neil Weinstock Netanel, *Copyright Paradox* (OUP 2008) 44; Lionel Bently and Brad Sherman, *Intellectual Property Law* (OUP 4th edn, 2014) 302-04.
[50] Lemley (n 1), 503.

*Steamed Bun*, a parody of Chen's movie that became fairly popular on the Internet.[51] Although some commentators claimed that the parody eventually helped promote Chen's original work, Chen insisted that it was a shameless case of copyright infringement.[52] A more recent case is that of a famous *wuxia* novelist, Jin Yong, who claimed that the young writer Jiang Nan, whose fiction borrowed characters from Jin Young's novels, had engaged in copyright infringement.[53] The Guangzhou District Court held that although Jiang Nan did not infringe Jin Young's copyright, the former's use of the characters in the latter's book violated the Anti-Unfair Competition Law.[54] Consequently, the court ruled that the defendant had to pay Jing Young damages in the amount of RMB$1,680,000.[55] These are examples in which reputable copyright owners adopt a zero-tolerance policy toward parties engaging in Type I copyright sharing.

By contrast, some copyright owners have shared their work via Creative Commons and free or open-source software licenses. In the music industry, copyright owners share their work either voluntarily or involuntarily, and they all aspire to monetise the sharing of their works. We discuss copyright-sharing activities originated by the copyright holders below.

### 3.1. Creative commons community

Creative Commons (CC), a set of flexible licensing terms that helps authors share their copyright, has been viewed as a major facilitator of the emergence of the copyright-sharing economy.[56] Copyright owners may choose different types of CC licenses in order to share their works under different conditions.[57] Because the application of CC represents an important part of the copyright-sharing economy, it is worthwhile to observe the development of CC in China.

---

[51] Han (n 16) 99; Robert S. Rogoyski1 and Kenneth Basin, 'The Bloody Case That Started from a Parody: American Intellectual Property and the Pursuit of Democratic Ideals in Modern China' (2009) 16 UCLA Ent L Rev 237, 239-40.

[52] '陈凯歌愤怒起诉《一个馒头引发的血案》(Chen Keigo Angrily Suing the Party Producing "The Bloody Case that Started from a Steamed Bun")' (*Tencent News*, 13 February 2006) <https://news.qq.com/a/20060213/000880.htm>, accessed 21 January 2019.

[53] '金庸告江南《此间》侵权索赔 500 万 江南:摸不着头脑(Jin Young Sues Jiang Nan for Copyright Infringement and Claims 5 Million Dollars Damage Whereas the Latter Doesn't Understand the Claim at All' (*Sina Entertainment*, 24 October 2016) <http://eladies.sina.com.cn/news/star/2016-10-24/1604/doc-ifxwztrt0252254.shtml> accessed 21 January 2019.

[54] 查某某訴楊某等著作權侵權及不正當競爭糾紛案 (Cha v Young), 廣東省廣州市天河區人民法院 (Guangzhou Tianhe District Peoples' Court), (2016)粤 0106 民初 12068 號 (2016 Yue 0106 Minchu No. 12068), 16 August 2018.

[55] Ibid.

[56] Benkler (n 9) 455-56; Joanna E Collins, 'User-Friendly Licensing for a User-Generated World: The Future of the Video-Content Market' (2013) 15 Vand J Ent & Tech L 407, 432-34; Séverine Dusollier, 'The Master's Tools v. the Master's House: Creative Commons v. Copyright' (2006) 29 Colum J L & Arts 271, 272-74; Farchy (n 14) 257-58; Johnson (n 7) 1978-80; Gideon Parchomovsky, Philip J. Weiser, 'Beyond Fair Use' (2010) 96 Cornell L Rev 91, 123.

[57] Lawrence Lessig, *Free Culture: How Big Media Uses Technology and the Law to lock Down Culture and Control Creativity* (The Penguin Press 2004) 282-86.

CC China was launched in 2006.[58] It has become quite successful in some scenarios, such as open education resources[59] and community photography.[60] CC has been adopted for massive open online courses (MOOC) platforms in China,[61] such as NetEase Open Courseware (https://open.163.com/) and Chinese Universities MOOC (http://www.core.org.cn/). Some scholars have licensed their academic works under CC as well.[62] Nevertheless, the application of CC in China has been quite limited—a limitation evident in other jurisdictions as well.[63] As the former project leader of CC Belgium Séverine Dusollier observed, CC licenses mostly interest creators whose 'primary purpose in creation might not be remuneration', and these creators 'might include the teachers, scientists, non-governmental organizations or even artists.'[64] After all, such creators constitute only a small portion of copyright owners.

The development dilemma faced by CC China is similar to that in other jurisdictions. After all, there has been no sustainable business model based on CC licenses, which are rarely seen in any major commercial website in the country. This is a common predicament for the sustainable development of the Type II sharing. Moreover, China lacks a sizable CC community. Moreover, China lacks a sizable CC community. Because authors' implementation of CC licenses is based on voluntary private ordering, it takes strong norms of sharing in the community for CC to reach a critical mass.[65] Nevertheless, the lack of both robust sharing norms and a sizable sharing community has become the main obstacle for CC's expansion to the next stage in China and many other jurisdictions. **\*439** Although CC provides an alternative means by which copyright owners can share their works, it may not align the interests of many copyright owners or, more importantly, of online platforms connecting creators and users.

---

[58] Creative Commons China, '宗旨和历史(Objective and History)' < http://creativecommons.net.cn/about/history/> accessed 21 January 2019.

[59] Creative Commons China, '2013CC 年大事记之"开放学校"( 2013 CC Memorabilia: Open School)' < http://creativecommons.net.cn/2013cckfxx/> (2014) accessed 21 January 2019.

[60] Creative Commons China, '观看中国—CC 摄影作品展 II 征稿启事 (Observing China: Call for Submissions of CC Photography Exhibition)' (2012) < http://creativecommons.net.cn/1-34/> accessed 21 January 2019; Wangyi Photography, '网易摄影知识共享说明 (Explanation Regarding Wangyi Photography Licensed under Creative Commons)' < http://pp.163.com/html/creativecommons> accessed 21 January 2019.

[61] NetEase, NetEase Open Courseware < https://open.163.com/> accessed 28 21 January 2019; Creative Commons China, '北京大学慕课工作组组长李晓明教授与 CC 中国大陆探讨合作 (Peking University MOOC team leader, Prof. Li Xiaoming, Exploring Collaboration with Creative Commons China' (2017) < http://creativecommons.net.cn/prof-li-of-pku-mooc-visits/> accessed 21 January 2019.

[62] China News Network 'First Medicine Book Licensed Under CC in Mainland China Being Published(中国内地首部使用知识共享协议出版的医学著作问世)' (*Science Net*, 2009) < http://news.sciencenet.cn/htmlnews/2009/7/221149.shtm> accessed 21 January 2019.

[63] Collins (n 56) 432-34.

[64] Dusollier (n 56) 282.

[65] Séverine Dusollier, 'Sharing Access to Intellectual Property Through Private Ordering' (2007) 82 Chi-Kent L Rev 1391, 1411-13; Amy Kapczynski, 'The Access to Knowledge Mobilization and the New Politics of Intellectual Property' (2008) 117 Yale LJ 804, 873.

## 3.2. Free or open-source software developers

Free or open-source software (F/OSS) is software with a source and object code that are distributed and made available to the public, allowing for free modification by other programmers. [66] In contrast, most commercial software is proprietary and distributed only with the object code to keep competitors from reusing the software for further development. [67] F/OSS is distributed under a license that requires source-code authors, distributors, and users to comply with certain conditions. [68] The F/OSS community has a fundamental notion and culture of innovation through sharing. [69]

China has numerous F/OSS communities, most of which are loosely organised. As of 17 January 2019, there were already 10,096 local F/OSS projects on OS China, a major F/OSS repository platform in the country. [70] Major Chinese technology companies, such as Alibaba, Tencent, and Huawei, have started to invest in F/OSS projects. [71] For example, Alibaba has initiated more than 150 F/OSS projects and was included in the world's leading F/OSS platform Github's list of top contributors in 2017. [72] Tengine is a sizable F/OSS project sponsored by Taobao and Sogou in the Alibaba group. [73] Both Huawei and Tencent actively participate in the community of OpenStack, an F/OSS cloud management platform, and share code and experiences with the community. [74] Huawei has invested and participated heavily in the Linux project as well. [75] Moreover, Baidu has open-sourced its machine-learning tools

---

[66] Boldrin and Levine (n 3) 20; Farchy (n 14) 255-56; Jyh-An Lee, 'New Perspectives on Public Goods Production: Policy Implications for Open Source Software' (2006) 9 Vand J Ent & Tech L 45, 49.

[67] Lee (n 66) 49.

[68] Ibid 50.

[69] Benkler (n 9) 63-67; David (n 3) 66; Dusollier (n 65) 1398-1400; Johnson (n 7) 1976-78; Lessig (n 57) 163-66, 174-75; Kal Raustiala and Christopher Springman, The Knockoff Economy: How Imitation Sparks Innovation (OUP 2012) 186; Farchy (n 14) 255; Guido Westkamp, 'The Limits of Open Source: Lawful User Rights, Exhaustion and Co-existence with Copyright Law' (2008) 1 IPQ 14, 23, 57.

[70] OSChina, 'Domestic Open Source Projects' <https://www.oschina.net/project/zh> accessed 21 January 2019.

[71] Linux Foundation, 'Tencent and Why Open Source is About to Explode in China' (19 February 2016) < https://www.linuxfoundation.org/blog/tencent-and-why-open-source-is-about-to-explode-in-china/> accessed 21 January 2019.

[72] Alibaba Tech, '12 Alibab Techs made Open-Source in 2017' < https://medium.com/@alitech_2017/alibabas-open-source-core-technologies-of-2017-2734ba5c154a> accessed 21 January 2019.

[73] Tengine, 'Introduction' <http://tengine.taobao.org/> accessed 21 January 2019.

[74] Nicole Martinelli, 'How Tech Giant Tencent Uses OpenStack' (*Superuser*, 2017) <http://superuser.openstack.org/articles/tencent-openstack/> accessed 21 January 2019; Jim Zemlin, 'Open Source Powers the Cloud Ecosystem' (2017) <http://www.huawei.com/en/about-huawei/publications/communicate/81/open-source-powers-cloud-ecosystem> accessed 21 January 2019.

[75] Linux Foundation, 'Huawei Deepens its Investment in Linux and Open Source Software with Linux Foundation Platinum Membership" (2015) < https://www.linuxfoundation.jp/press-release/huawei-deepens-its-investment-in-linux-and-open-source-software-with-linux-foundation-platinum-membership/> accessed 21 January 2019.

PaddlePaddle [76] and the deep-learning project DeepBench, which has attracted contributions from Intel, Nvidia, and AMD. [77] F/OSS is also rooted in student communities. Both the Linux User Group at the University of Science and Technology of China [78] and the TUNA Association at Tsinghua University [79] are active student F/OSS communities. Such student communities have emerged because F/OSS development has provided students with extensive resources with which to learn coding and programming for free. [80]

F/OSS developers contribute code without any monetary compensation for various reasons. Some believe the F/OSS community is a gift culture motivated by altruism and reciprocity; [81] others suggest that programmers aim to signal their abilities to others by solving complex problems or contributing significant new pieces of software to the community. [82] Such signalling brings about a feeling of satisfaction [83] and reputability that eventually benefits programmers' careers. [84] In other words, Type II sharing and Type III sharing coexist in the F/OSS development. Indeed, some Chinese F/OSS contributors view their sharing and participation as a strategy for seeking positions in large companies. [85] Nevertheless, **\*440** empirical studies have suggested that compared to their counterparts in India, Chinese F/OSS developers are driven mostly by intrinsic motives, such as personal needs, reputation, skill-gaining benefits, and enjoyment of coding. [86] Regardless of why Chinese F/OSS developers contribute their talents, given the increasingly robust communities and businesses' continuous investment in this field, F/OSS has become one of the most representative examples of voluntary copyright sharing in the country.

---

[76] Alex Contini, 'OpenDaylight Project Expands in China with Baidu' (2016) < https://www.opendaylight.org/foundation-news/2016/12/08/opendaylight-project-expands-in-china-with-baidu-2> accessed 21 January 2019; IDG Connect, 'Baidu Open Sources Its Deep Learning Platform PaddlePaddle' (2017) <http://www.idgconnect.com/abstract/20079/baidu-sources-deep-learning-platform-paddlepaddle> accessed 21 January 2019.

[77] A.R. Guess, 'Baidu Research Announces Next Generation Open Source Deep Learning Benchmark Tool' (*Dataversity*, 2017) <http://www.dataversity.net/baidu-research-announces-next-generation-open-source-deep-learning-benchmark-tool/> accessed 21 January 2019.

[78] Linux User Group (LUG) in the University of Science and Technology of China (USTC), 'Linux User Group at USTC' <https://lug.ustc.edu.cn/wiki/intro_english> accessed 21 January 2019.

[79] TUNA Association in Tsinghua University, TUNA Association in Tsinghua University, <https://www.tuna.moe/> accessed 21 January 2019.

[80] Eric von Hippel and Georg von Krogh, 'Open Source Software and the "Private-Collective" Innovation Model: Issues for Organization Science' (2003) 14 Organization Science 209, 216-19.

[81] Klaus M Schmidt and Monika Schnitzer, 'Public Subsidies for Open Source? Some Economic Policy Issues of the Software Market' (2003) 16 Harv JL & Tech 473, 481.

[82] Lerner and Tirole (n 34) 214.

[83] Steven Weber, *The Success of Open Source* (Harvard University Press 2004) 140.

[84] James Bessen, 'What Good Is Free Software?' in Robert H Hahn (eds), *Government Policy Toward Open Source Software* (Brookings Institution Press 2002) 12, 18.

[85] Tong Hui, '两岸开源社群面面观 (Various Aspects on Open Source Communities Crossing the Strait)' (*Business Next*, 2016) <https://www.bnext.com.tw/article/38447/BN-2016-01-11-000328-77> accessed 21 January 2019.

[86] Kevin Crownston, Kangning Wei, James Howison, and Andrea Wiggins, 'Free/Bibre Open-Source Software Development: What We Know and What We Do Not Know' (2012) 44(2) ACM Computing Surveys 7:1, 7:13.

Although F/OSS development has achieved preliminary success in China, it still falls behind that in the Western world in terms of both quality and quantity. One possible explanation is that ineffective IP protection and enforcement have negatively affected F/OSS development in China.[87] This explanation may not sound plausible, because the ideology underlying the F/OSS movement is a reaction to the overly expansive IP exclusion. However, the entirety of F/OSS development is in fact still based on IP.[88] Without an effectively operating IP regime, F/OSS licenses, which are the foundation of F/OSS development, cannot function at all. Therefore, the case of F/OSS shows that the copyright-sharing economy is not essentially in conflict with the IP system. Moreover, it should be noted that the enforceability of F/OSS licenses and CC licenses have both been recognized by the Chinese courts. In 2018 the Beijing Intellectual Property Court held that the GNU Public License (GPL), which is the most common F/OSS license requiring every derivative work be licensed under the same GPL, is enforceable.[89] The Ningbo Intermediate People's Court in Zhejiang similarly ruled that both F/OSS and CC licenses were enforceable.[90] The Court further explained that the purpose of these open licenses is to enhance social welfare by sharing and allowing users to exploit the underlying IP without any licensing fee.[91] The Court held that the fact a user decides to use a specific piece of F/OSS means that he accepts the licensing terms attached to the software.[92] Therefore, the legality of open licenses, which institutionalize the community norm of sharing, has been acknowledged by the Chinese judiciary. These line of court decisions have created a friendly legal environment for the CC and F/OSS development in China.

Compared to CC, F/OSS communities in China seem to be a more sustainable sharing model because of the unique process of software creation and businesses' aligned interests in the code-sharing economy. Such a distinction exists not only in China but also elsewhere in the world. Some commentators have argued that this is the case because software belongs to a unique category of creation in which perfection and optimal performance cannot be overemphasised.[93] The open sharing and collaborative nature of F/OSS rightfully responds to this demand. By contrast, authors of most cultural creations care about the integrity of their works, and elimination of errors is not an essential part of their learning and creation.[94] As a result, sharing and collaboration are not the best mode for creativity. In addition, the F/OSS communities in China and other countries have successfully developed an ecosystem in which commercial entities 'leverage value from a sharing economy' or a sharing economy builds commercial entities 'to better support its commercial aim'.[95] Lawrence Lessig referred to this ecosystem as a 'hybrid economy'.[96] Put differently, businesses' incessant investment in F/OSS has facilitated the continued development of this sharing economy.[97] Type II sharing therefore has thrived because of the vigorous development of Type III sharing. By contrast, the comparative slowdown of CC adoption in China might be due to an insufficient incentive for businesses to share their content under CC licenses. A economically viable institutional model of Type II sharing for CC has yet to be developed in China. This may explain why the sharing economy is more robust and sustainable in F/OSS than in other forms of cultural creativity.

---

[87] Yongming Wei, '开源软件及国内发展现状 (Open Source Software and Its Current Development in China)' (Open Source China, 2012) <https://www.oschina.net/news/33260/china-opensource-status> accessed 21 January 2019.
[88] Lee (n 66) 50.
[89] Digital Heaven Ltd v APICloud, Jing Zhi Min Chu Zi No. 631 (Beijing Intellectual Property Court 2015).
[90] Zhejiang Avanti E-Commerce Ltd v Wangcheng Technology Co. Ltd., Zhe 02 Min Chung No. 3852 (Ningbo Intermediate People's Court 2017).
[91] Ibid.
[92] Ibid.
[93] Farchy (n 14) 259-60.

### 3.3. The music industry

Some copyright researchers have argued that the music industry's business models have substantially changed the form of the sharing economy.[98] Artists are now willing to share their copyright works to create reputational value.[99] Their profits come from live performances and merchandise instead of the sale of albums.[100] This is because albums are easy to copy but the experience of a concert is not.[101] Some scholars have therefore argued that sharing music online is the best way for artists to be compensated[102] and that it thus represents a superior incentive for music creation.[103] This is a typical Type III sharing and the revenue model of Radiohead, Nine Inch Nails, Coldplay, and the Arctic Monkeys.[104]

In China, the music industry has developed in a much more complicated way than the music-sharing economy described above. Like their counterparts in other countries, artists in China seek a variety of alternative revenue streams, such as live performances, sponsorship, advertisements, merchandise, and synchronisation with movies, TV shows, video **\*441** games, and advertisements.[105] However, empirical evidence suggests that the shift of revenue streams and music business models in China is an attempt to protect the music industry from pervasive piracy.[106] The inescapable online sharing of music is not a decision made by copyright owners to build their reputation. Instead, this type of sharing economy is mostly associated with music piracy and has brought about huge challenges for the music industry.[107] Put differently, music sharing on a mass scale occurs mainly because copyright enforcement is increasingly expensive and relatively ineffective in this context. This looks more like a twisted version of the Type III sharing.

---

[94] Ibid.
[95] Lessig (n 57) 177.
[96] Ibid., 177-78.
[97] Ibid., 179-85; Raustiala and Springman (n 69) 188-89.
[98] David (n 3) 41; Giancarlo F. Frosio, 'User Patronage: The Return of the Gift in the "Crowd Society"' (2015) 2015 Mich St L Rev 1983, 2032.
[99] Frosio (n 98) 2032.
[100] Ibid; David (n 3) 53, 175.
[101] Raustiala and Springman (n 69) 222.
[102] David (n 3) 49.
[103] Ibid., 165.
[104] Ibid., 1; Frosio (n 98) 2033.
[105] Anderson (n 15) 199-201; Jiarui Liu, 'The Tough Reality of Copyright Piracy: A Case Study of the Music Industry in China' (2010) 27 Cardozo Arts & Ent LJ 621, 635-62.

Second, music copyright is still generating revenue from ringback tones provided by the telecom companies.[108] Ringback tones have been a much easier revenue source because they are integrated with mobile phone services and are centrally controlled by mobile service providers.[109] However, around 98% of this revenue has gone to three major telecom companies, namely China Mobile, China Unicom, and China Telecom, whereas only approximately 2% of the revenue has gone to copyright owners.[110] Chinese mobile providers' annual gross from the ringback is over US$4 billion.[111] If the music copyright owner had an equal share of this enormous revenue with the telecom service providers, China would become the number three country in the globe that generates most revenue for recorded-music copyright owners, only next to the United States and Japan.[112] To a certain degree, this phenomenon reflects the unbalanced bargaining power and consequent inequitable revenue distribution between telecom companies and the music industry with respect to music ringback-tone services.

Third, China's recording industry has developed a music-sharing business model in music streaming, which has become an increasingly important revenue source for the industry.[113] Music labels share music with users for free via streaming services, and the labels' main income derives from displaying ads in the streaming services.[114] Users are required to pay only if they desire higher sound quality, exclusive content, or free download services.[115] The 'freemium' music business model also exists in countries other than China. In the United States, Magnatune shares MP3 quality music with its users.[116] If the users purchase the album, they can enjoy high-quality music identical to that on a CD.[117] After all, music and other information goods are 'experienced goods'.[118] It is quite common for producers to share free samples with consumers, who can experience the music before they decide whether they are willing to pay for it.[119] Even a radio broadcast of a song can be viewed as an ad or free sample that is shared in an inconvenient or inferior-quality form.[120] If you want to enhance your enjoyment of that song, you will need to pay for it. Some commentators have maintained that the Type III freemium sharing model has become more sustainable with increased IP enforcement, and an increasing number of Chinese users are willing to pay for music streaming.[121] Nonetheless, most music labels in China are still struggling to turn nonpaying users into paying users.[122] This is the common challenge for all Type III sharing models.

---

[106] Ibid.; Priest (n 2) 1081; Raustiala and Springman (n 69) 182-84.
[107] Marco Figliomeni, 'The Song Remains the Same: Preserving the First Sale Doctrine for a Secondary Market of Digital Music' (2014) 12 Can J L & Tech 219, 236.
[108] Priest (n 15) 501-02.
[109] Montgomery & Priest (n 15) 353.
[110] Priest (n 2) 1086; Tang and Lyons (n 15) 358, 362.
[111] Montgomery & Priest (n 15) 353.
[112] Ibid 353.
[113] Tang and Lyons (n 15) 360.
[114] Ibid 360, 363.
[115] Ibid 360-61.
[116] Farchy (n 14) 262.
[117] Ibid.
[118] Shapiro & Varian (n 39) 85.

Fourth, with the advancement of digital technologies, musicians do not need to rely on expensive studios to record their music and big companies to distribute it.[123] Instead, they can easily produce high-quality music with their laptop and share it via the Internet.[124] Quite a few artists in the Western world, such as Lili Allen, Kate Nash, and Arctic Monkeys in the United Kingdom and Colbie Caillat, Mac Miller, and J. Cole in the United States, built their reputations as pop singers by sharing their works online and were eventually signed by major music labels.[125] This new model also belongs to the Type III sharing and has established a number of popular singers in China. For example, Silence Wang (汪苏泷), Liang Xu (徐良), and Vae Xu (许嵩) became famous singers by sharing their self-created songs online.[126] All of them eventually released their own albums after becoming Internet celebrities.

Digital technologies have reshaped the landscape of the music industry, and a new music-sharing economy has emerged in China. Unknown artists may quickly become popular singers by sharing their work online. Established music labels may induce consumers to pay for the better music experience of a live performance or for high-quality music by sharing music in an inferior-quality form. We have seen an institutional change in the use of copyright in this context. As Nobel laureate Douglas C. North indicated, a change in relative price will lead to a change in contractual parties'

---

[119] Ibid.

[120] Ibid 87.

[121] Tang and Lyons (n 15) 361; The International Federation of the Phonographic Industry, 'Focus on China: China's Phenomenal Potential Unlocked by Streaming' in *Global Music Report 2017*(2017); 'A Pirate's Life No More: Thanks to Streaming Services, China's Consumers have Begun Paying for Music' (2017) Economist < https://www.economist.com/news/21725529-market-dominated-one-company-tencent-biggest-countrys-online-giants > accessed 21 January 2019; 'Digital Music Fuels Development of China's Music Industry' (*iResearch*, 2018) < http://www.iresearchchina.com/content/details7_41344.html> accessed 21 January 2019.

[122] Tang and Lyons (n 15) 363-64.

[123] Lemley (n 1) 470; Raustiala and Springman (n 106) 214.

[124] Raustiala and Springman (n 69) 215.

[125] Ibid., 223-24.

[126] Bo Linlin, '许嵩徐良汪苏泷 时代造就的 "QQ 音乐三巨头" (Vae Xu, Liang Xu, and Silence Wang: Three QQ Music Giant Created by Our Time)' (QQ Entertainment) <http://ent.qq.com/a/20160521/012781.htm> accessed 21 January 2019.

**\*442**  behaviours.[127] Consequently, the tradition may be eroded and replaced by new practices.[128] In the case of the music industry in China and elsewhere, because the relative price of disseminating and reproducing music has dropped significantly, copyright owners need to develop new ways of using copyright. Type III sharing has created unique assets for copyright owners which may be monetized beyond traditional way of proprietarily exploiting copyright. As a result, the exchange value of copyright has shifted from IP exclusion to enticement by sharing in the music industry.

### 3.4. Summing up

The exclusive and nonexclusive use of IP will continue to coexist in the content industry.[129] Chinese copyright owners' diverse attitudes toward sharing their works illustrate the dynamic incentive structure in the copyright regime. Monetary incentives certainly continue to play a vital role in most of the country's cultural creation. Traditional authors are typically reluctant to share their copyright, whereas certain creative communities, such as educators and academics, are more willing to share their copyright via CC licenses or other Type II sharing mechanisms. Nonetheless, with the rapid development of digital and network technologies, Type III copyright sharing has become a new business strategy for copyright owners' creation of fame or extraction of commercial value from their works. Although IP sharing has been viewed as a nonmarket form of exchange,[130] copyright sharing in the F/OSS development and the music industry is at least partly driven by market forces. Businesses strategically leveraging F/OSS development or sharing their musical works as described above exemplify unconventional ways of sharing copyright for the sake of competitive advantage.

## 4. Copyright users

Copyright users typically embrace the sharing economy without reservation.  At the beginning of the digital era, users' large-scale sharing of musical works over the p2p network (Type I sharing) was the main target of copyright litigations concerning infringement.[131] Nevertheless, on the positive side, with the advancement of digital technologies, individuals have been able to make use of existing copyright works in novel ways and to participate in the cultural creations. Users of copyright works are not the couch potatoes of the pre-Internet era but authors who create based on existing works. The line between consumers and producers of content has been blurred in the digital environment, because consumers are occasionally producers too.[132] Authors who contribute their works to the aforementioned CC or F/OSS communities are

---

typically users of copyright works in the community as well. However, the legal restrictions on making use of existing copyright works have not been loosened in response to the advancement of digital technologies. Users who share and adapt existing works are subject to a high risk of infringement liabilities. It is quite common that such users share and remix others' work without obtaining permissions and consequently infringe others' copyright.[133] Although user-generated content (UGC) has become a crucial part of the copyright-sharing economy,[134] social media users may not be aware of the infringement risk associated with their use of other people's copyrighted work.[135] Under the Chinese Copyright Law (CCL), chances are quite high that users will infringe the right of reproduction[136] or the right of adaptation by conducting the Type I sharing.[137]

The copyright regime has a range of limitations or exceptions to copyright protection that are important mechanisms for facilitating a balance between the interests of copyright owners and of users.[138] These limitations are of great importance to the scope within which users can use and share existing copyrighted work.[139] Article 13 of the Agreement on Trade-Related Aspects of Intellectual Property Rights incorporates Berne Convention Article 9(2)'s three-step test for limitations or exceptions to exclusive rights by stipulating that 'members shall confine limitations or exceptions to exclusive rights to certain special cases which do not conflict with a normal exploitation of the work and do not unreasonably prejudice the legitimate interests of the right holder'.[140] Based on the three-step test, limitations or exceptions take different forms at the level of national law.[141] In most commonwealth jurisdictions, which adopt the 'fair dealing' approach, and most civil law jurisdictions, copyright laws provide a limited list of exceptions to copyright, which typically include uses for the **\*443** purposes of research, education, and media reporting.[142] Defendants or users must defend their uses based on at least one of these specified exceptions.[143] By contrast, Section 107 of the U.S. Copyright Act of 1976 takes an open-ended approach to copyright exception by establishing four factors used by courts to determine fair use: (1) the purpose and character of the use, including whether the use is of a commercial nature or is for nonprofit educational purposes, (2) the nature of the copyrighted work, (3) the amount and substantiality of the portion used in relation to the copyrighted work as a whole, and (4) the effect of the use upon the potential market for or value of the copyrighted work.[144] Courts may also consider other factors when weighing a fair-use question.

---

[133] Bertoni and Montagnani (n 134) 398; Collins (n 56) 411.

[134] Alm (n 48) 108; Aura Bertoni and Maria Lillà Montagnani, 'Foodporn: Experience-sharing Platforms and UGC: How to Make Copyright Fit for the Sharing Economy' (2017) 39(7) EIPR 396, 397; Montgomery & Priest (n 15) 342.

[135] Alm (n 48) 117-20; Kanchana Kariyawasam and Chamila Talagala, 'Cloud Computing and the Future of Copyright Law' (2017) 39(2) EIPR 102, 105; Monica Yun, 'Pinterest's Secondary Liability: The DMCA Implications of Holding Pinterest Responsible and What Pinterest Can Do to Avoid Liability' (2013) 36 Hastings Comm & Ent LJ 489, 490.

[136] 2010 Copyright Law of the People's Republic of China, art. 10(5).

[137] Ibid. art. 10(3).

[138] Tanya Aplin and Jennifer Davis, *Intellectual Property Law: Text, Cases, and Materials* (3rd edn, OUP 2017) 222-23; Tatiana-Eleni Synodinou, 'The Lawful User and A Balancing of Interests in European Copyright Law' (2010) 41(7) IIC 819, 819.

[139] Reto M. Hilty, 'European Commission: Green Paper: Copyright in the Knowledge Economy - Comments by the Max Planck Institute for Intellectual Property, Competition and Tax Law' (2009) 40(3) IIC 309, 317-18; Helena R. Howe, 'Copyright Limitations and the Stewardship Model of Property' (2011) 2 IPQ 183, 198; Waelde and MacQueen (n 5) 282.

[140] TRIPs Agreement Article 13. The three steps include: (1) "certain special cases," (2) "which do not conflict with a normal exploitation of the work" and (3) "[which] do not unreasonably prejudice the legitimate interests of the right holder."

[141] Paul Goldstein and Bernt Hugenholtz, *International Copyright: Principles, Laws, and Practices* (3rd edn, CUP 2013) 377.

Although in Article 22 of the CCL there is a limitation and exception clause, which is typically refereed to fair use, the chances for copyright users to apply this defence were formerly quite slim.[145] This is because Article 22 provides only 12 types of exceptions and limitations to copyright protection, such as 'for the purposes of the user's own private study, research or self-entertainment',[146] 'for the purposes of introduction to, or comments on, a work, or demonstration of a point',[147] and 'translation, or reproduction in a small quantity of copies, of a published work for use by teachers or scientific researchers, in classroom teaching or scientific research, provided that the translation or reproduction shall not be published or distributed'.[148] From a statutory perspective, use of a copyright work that fails to fall into any of the 12 categories can hardly be viewed as fair use.[149] Therefore, some commentators have suggested that 'free use', instead of 'fair use', should be used to refer to the list of exceptions in Article 22 of the CCL.[150] To accommodate fair use in the digital environment, Article 6 of the Regulations for the Protection of the Right of Communication through the Information Network (the 2016 Information Network Regulations) supplements the 12 exceptions listed in the CCL's Article 22.[151]

From a comparative law perspective, the fair-use clause in the U.S. Copyright Act provides more flexibility than that in the CCL. Compared to the Chinese approach, the U.S. fair-use approach provides more space for users to defend against infringement claims, because the fair-use clause in the CCL provides only a closed list of exceptions. For example, users who create parodies in the United States may claim fair use for their works,[152] whereas such users in China may find it difficult to receive exemptions from infringement liability.[153] The fair-use regime in the CCL has long been criticised as overly restrictive and lacking flexibility.[154] In this regard, the CCL creates a less friendly environment for copyright users and the sharing economy. Nevertheless, the flaw of the U.S. fair-use approach is that it may create more uncertainties for users than China's closed-list approach.[155]

---

[142] Aplin and David (n 138) 224; Goldstein and Hugenholtz (n 141) 375-76.
[143] Aplin and David (n 138) 224-25.
[144] 17 U.S.C. § 107.
[145] Tianxiang He, *Copyright and Fan Productivity in China: A Cross-jurisdictional Perspective* (Springer 2017) 128-130, 135-37; Han (n 16) 103.
[146] 2010 Copyright Law of the People's Republic of China, art. 22(1).
[147] Ibid. art. 22(2).
[148] Ibid. art. 22(6).
[149] Li and Greenleaf (n 19) 164; (Jerry) Jie Hua, 'Incorporation of Incidental Use into Copyright Limitations and Exceptions in China' (2017) 39 EIPR 30, 30.
[150] Li and Greenleaf (n 19) 164-68; Yong Wan, 'Similar Facts, Different Outcomes: A Comparative Study of the Google Books Project Case in China and the United States' (2016) 63 J Copyright Soc'y USA 573, 575.
[151] Regulations for the Protection of the Right of Communication through the Information Network (promulgated by the State Council, 10 May 2006) art. 6.

In recent years, the Chinese judiciary has acknowledged the downside of its copyright exception in Article 22 of the CCL and moved gradually toward a U.S.-style nonexclusive fair-use approach. For example, the Beijing Haidian District Court found that the China Education TV Station's use of SARFT Movie Center without permission was not infringement.[156] The court first confirmed that the defendant's use did not fall into any of the categories enumerated in Article 22, and then applied the two-factor test to find that it was fair use.[157] These two factors are the first and last factors of the U.S. fair-use model: (1) the purpose of the use, and (2) the effect of the use on the market.[158] In 2007, the Shandong High Court held that the defendant's use was fair based on three factors: (1) the purpose of the use, (2) the amount of the use, and (3) the effect of the use on the market.[159] The three factors considered by the court were obviously transplanted from the four factors in Section 107 of the U.S. Copyright Act.[160]

In 2012, the Chinese Supreme Court issued a policy document recognising the possibility of implementing the U.S. fair-use approach:

> In the definitely necessary circumstances to stimulate
> technical innovation and commercial development, an act
> that would neither conflict with the normal exploitation
> of the work nor unreasonably prejudice the legitimate
> interests of the author, may be considered free use,
> provided that the purpose and character of the use of the

---

[152] Campbell v. Acuff-Rose Music, Inc. 510 U.S. 569 (1994).
[153] Rogoyski1 and Basin (n 51) 243-44; Haochen Sun, 'Can Louis Vuitton Dance with Hiphone? Rethinking the Idea of Social Justice in Intellectual Property Law' (2012) 15 U Pa J L & Soc Change 389, 429-30.
[154] Seagull Haiyan Song, 'Reevaluating Fair Use in China—A Comparative Copyright Analysis of Chinese Fair Use Legislation, the U.S. Fair Use Doctrine, and the European Fair Dealing Model' (2011) 51 IDEA 453, 453, 480.
[155] Benkler (n 9) 440-41; Netanel (n 49) 66; Pierre N. Leval, 'Toward a Fair Use Standard' (1990) 103 Harv L Rev 1105, 1106-07; David Nimmer, '"Fairest of Them All" and other Fairy Tales of Fair Use' (2003) 66 Law & Contemp Probs 263, 267; Michael J. Madison, 'Rewriting Fair Use and the Future of Copyright Reform' (2005) 23 Cardozo Arts & Ent LJ 391, 402.
[156] Song (n 154) 483.
[157] Ibid.
[158] Ibid.
[159] Song (n 154) 481-82.
[160] 17 U.S.C. § 107.

> work, nature of the work, amount and substantiality of the
> portion taken, and effect of the use upon the potential
> market and value have been taken into account.

The Supreme Court first identified two elements of the three-step test as the precondition of fair use, namely (1) not *444 to conflict with the normal exploitation of the work, and (2) not to unreasonably prejudice the legitimate interests of the author. It then embraced the four factors of U.S. fair use entirely.

In a highly controversial 2013 case regarding the Google Book Project, the Beijing Higher Court adopted a fair-use standard similar to that in the aforementioned policy document.[161] In that case, the plaintiff, Xin Wang, a famous novelist widely known by her pseudonym Mian, claimed that Google infringed her copyright over the novel *Yansuan Qingren* (Acid Lovers).[162] Google borrowed a copy of the book from Stanford University Library and scanned it in its entirety in the United States without the plaintiff's permission.[163] As with other books included in the Google Book Project, users can employ search words to access relevant snippets from the book.[164] Google's defence was that its use was fair use, not infringement.[165] When evaluating Google's fair-use defence, the Beijing Higher Court ruled that fair use does not necessarily need to fall into the specified categories listed in Article 22 of the CCL;[166] instead, under some exceptional circumstances, the court should consider all relevant factors, including (1) the purpose and character of the use, (2) the nature of the copyrighted work, (3) the character and amount of the portion used in relation to the copyrighted work as a whole, (4) whether such use affected the normal exploitation of the work, and (5) whether such use unreasonably prejudiced the copyright holder's legitimate interests.[167] Moreover, users bear the burden of proof regarding the factors supporting fair use.[168]

In summary, the Beijing Higher Court introduced a multifactor test to determine fair use. Among the five factors presented by the court, three of the four U.S. fair-use factors were adopted.[169] The other two factors were taken from the three-step test. This approach, combining the three-step test and the U.S. four factors, resembles that suggested in the Supreme Court's 2012 policy document mentioned above. Both the Supreme Court and the Beijing Higher Court omitted the first step of the three-step test, 'limited to certain special cases.' Although the Beijing Higher Court implemented a multifactor test, which is similar to the U.S. law, to determine fair use, it reached a very different conclusion from the U.S. court decision. In the United States, the Second Circuit affirmed the district-court decision that the Google Book Project's copying and making snippers available to users was 'transformative' and therefore satisfied the test for fair use.[170] By contrast, the Beijing Higher Court eventually held that Google failed to provide sufficient evidence that its use constituted exceptional circumstances as fair use.[171] Although the Beijing Higher Court ruled against the copyright user (Google) in this case, it has undoubtedly crafted a wider range of fair use than that defined by Article 22 in the CCL. Some commentators have argued that as a result, Google might have won this case if it had provided sufficient evidence to prove its fair use.[172]

---

[161] *Google, Inc. v. Xin Wang*, No. 1221 Gaominzhongzi (Beijing Higher Ct. 2013).
[162] Ibid.
[163] Ibid.
[164] Ibid.
[165] Ibid.
[166] Ibid.
[167] Ibid.
[168] Ibid.
[169] Ibid.

If Chinese courts continue to rule based on the open-ended approach to resolve cases outside the specific categories in Article 22 of the CCL, they will likely shape a much friendlier legal environment for the copyright-sharing economy. Furthermore, it should be noted that in the Third Draft Amendment of the Copyright Law ('Draft') released by the National Copyright Administration of China (NCAC) in October 2012, an open clause was added to the copyright exceptions. According to Article 42.1 of the Draft, users can freely use others' work without the permission of copyright owners 'under other circumstances', and the interpretation of such other circumstances is based on Article 42.2, which states that such use will 'neither conflict with the normal exploitation of the work nor unreasonably prejudice the legitimate interests of the author'.[173] Like the 2012 Supreme Court document and the Beijing Higher Court's Google decision, the Draft incorporates the three-step test. Nevertheless, it does not include any of the four factors from the U.S. fair-use doctrine. Some scholars view the Draft as "a hybrid model that combines an enumerated list of permissible uses with a multifactor analysis."[174] The Draft obviously creates more space for users in the sharing economy, but a clearer standard for implementing the three-step test in this open exception would still need to be defined by the law or court decisions were the Draft adopted. No matter whether the Draft will be passed as its current version, it is clear that the scope of fair use, or limitation and exception of copyright law, will significantly influence users' incentive to conduct Type I sharing.

## 5. Online platforms

Online platforms, such as Google, YouTube, Facebook, and Flickr, have played an important role in facilitating the copyright-sharing economy.[175] While copyright

---

[170] *Authors Guild v. Google, Inc.*, 804 F.3d 202, 230 (2d Cir. 2015).
[171] *Google, Inc. v. Shen Wang* (n 161).
[172] Wan (n 150) 593-94.
[173] National Copyright Administration of the P. R. China, Third Draft Amendment of the Copyright Law, art. 42 (October 2012) <http://www.fengxiaoqingip.com/law/lawzz/jzqfl/20150302/10124.html> accessed 21 January 2019.
[174] Merges and Song (n 27) 456.
[175] Alm (n 48) 115-17; David (n3) 78-82, 156, 172; Brian Fitzgerald, 'Copyright in the Age of Access' (2017) 39 EIPR 131, 133; Lemley (n 1) 485-86; Brette G. Meyers, 'Filtering Systems or Fair Use? A

users occasionally enjoy the content shared via the platforms, copyright owners normally view those platforms as gatekeepers for copyright protection.[176] Nonetheless, the history of Internet commerce has witnessed the failure of quite a few online platforms because of copyright infringement liability. **\*445** Notable examples include the file-sharing websites Napster,[177] Grokster,[178] etc. The copyright practices of online platforms reflect their power dialectics with stakeholders, such as copyright owners, contributors, and users.[179] Such practices have a significant impact on the copyright-sharing economy. In China, online platforms have been blamed for unauthorized copyright sharing (Type I sharing) and widespread infringing activities.[180] Nonetheless, in recent years these platforms have developed more sophisticated business models facilitating legal sharing of copyright.

### 5.1 Notice-and-takedown procedure

As in many other jurisdictions, in China online platforms are subject to copyright secondary liability for their users' infringing behaviour, and such liability can be exempted in the safe-harbour regime. Article 36 of the Tort Liability Law stipulates the liability of Internet service providers.[181] The 2006 Information Network Regulations include a set of safe-harbour provisions.[182] In order to avert secondary liability, platforms are encouraged to implement notice-and-takedown procedures to deal with alleged infringing content uploaded by their users.[183] The Regulations provide guidelines for the interaction of online platforms with copyright owners and users with respect to possible infringing materials on their websites.[184] The purpose of the notice-and-takedown procedure is to incentivize platforms to curb the Type I sharing. Nonetheless, platforms' notice-and-takedown practices change over time and differ from platform to platform.

In 2009, the Chinese search-engine giant Baidu launched its Baidu Library project, widely referred to as Baidu Wenku in Chinese.[185] The project facilitated Type I sharing by allowing users to share, search, browse, and download books for free.[186] Baidu Library had stockpiled 2.8 million book files, most of which were still protected

---

Comparative Analysis of Proposed Regulations for User-Generated Content' (2009) 26 Cardozo Arts & Ent LJ 935, 935-36, 939; Yun (n 135) 489-90.

[176] Bonadio (n 23) 621.

[177] *A&M Records, Inc. v. Napster, Inc.*, 239 F.3d 1004 (2001).

[178] *MGM Studios, Inc. v. Grokster, Ltd.*, 545 U.S. 913 (2005).

[179] Guy Pessach, 'Some Realism about Copyright Skepticism' (2017) 57 IDEA 227, 269.

[180] Montgomery & Priest (n 15) 343.

[181] Tort Liability Law of the People's Republic of China, art. 36.

[182] Regulations for the Protection of the Right of Communication through the Information Network (n 151) art. 14-17, 20-25.

[183] Ibid.

[184] Ibid. art.14-17.

[185] AFP, 'Baidu Apologies to Writers in Copyright Dispute' < https://phys.org/news/2011-03-baidu-apologises-writers-copyright-dispute.html> accessed 21 January 2019.

[186] Ibid.

by copyright law at the time.[187] Although quite a few copyright owners notified Baidu to take down the files of their copyrighted works, Baidu neither responded nor immediately took down the controversial content. The project drew serious criticisms and complaints from a significant number of authors and publishers.[188] In addition to being forced to apologise to copyright owners,[189] Baidu was eventually held liable for copyright infringement by the court[190] and fined by the NCAC.[191]

Some commentators have maintained that Baidu project is quite similar to the Google Book Project.[192] Although Google and Baidu both engaged in Type I sharing in their respective projects, in contrast to Baidu's UGC book sharing service, Google itself scanned books from several academic and public libraries into its database. Google users were able to read the full text of public-domain materials and some snippets of books with copyright protection. Books in the Baidu Library, by contrast, were not scanned by Baidu itself; they were instead uploaded by users. However, like the Baidu Library project, the Google Book Project took heat from copyright owners. Google successfully relied on the fair-use defence for the reproduction of copyrighted works for archival and retrieval purposes, as long as only snippets from the copyrighted works were made available to the public.[193]

The Baidu Library case has several implications for the copyright-sharing economy in China. First, there was a period of time in which Chinese online platforms focusing on new business models based on Type I sharing of UGC exhibited insufficient awareness of the legal risk associated with copyright infringement. It is well noted that these platforms were notorious for gaming and delaying the notice-and-takedown process.[194] Baidu, for instance, for years had been alleged largest infringer of copyright, specifically music copyright, in the country.[195] Nonetheless, major platforms have started to enforce the notice-and-takedown procedure in a more effective way and to purge infringing content since 2009. In other words, their copyright strategies have disfavoured Type I sharing since then. Second, although there are noticeable differences between the models of the Baidu Library and the Google Book Project, Google's success in defending its fair use once again illustrates the U.S. Copyright Act's open approach to limitations and exceptions to copyright, which may provide a friendlier environment for platforms and new business models in the sharing economy. This illustrates the scope of copyright limitation and exception plays a critically important role in shaping Type I sharing.

---

[187] Loretta Chao, 'Baidu Takes Authors' Fire' (*Wall Street Journal*, 29 March 2011) < https://www.wsj.com/articles/SB10001424052748703739204576228532574791862> accessed 21 January 2019.
[188] Ibid.
[189] Ibid.
[190] You Yunting, 'Baidu Library Makes Huge Compensation on Copyright Infringement for Publishing Press' (*China IP Law Commentary*, 20 March 2014) < http://www.chinaiplawyer.com/baidu-library-make-huge-compensation-copyright-infringement-publishing-press/> accessed 21 January 2019.
[191] Patrick Brzeski, 'Chinese Search Giant Baidu Fined for Copyright Infringement' (*Hollywood Reporter*, 1 January 2114) < https://www.hollywoodreporter.com/news/chinese-search-giant-baidu-fined-668155> accessed 21 January 2019.
[192] Emily Gische, 'Repercussions of China's High-tech Rise: Protection and Enforcement of Intellectual Property Rights in China' (2012) 63 Hastings LJ 1393, 1409.
[193] *Authors Guild v. Google, Inc.*, 804 F.3d 202, 212-29 (2d. Cir. 2015).
[194] Montgomery & Priest (n 15) 347.
[195] Ibid 353.

Notice-and-takedown policy has become a rather common and effective practice among online platforms in China in recent years. Although some popular video-sharing platforms such as Bilibili (https://www.bilibili.com/) have occasionally become involved in copyright disputes over content generated *446 by users,[196] in general, copyright awareness has increased considerably. This awareness may partly be the result of an increasing number of lawsuits brought by copyright owners against online platforms.[197] Moreover, major video-sharing website, such as Youku, Tudou, and Sohu, have developed new business models and avoided facilitating users to share infringing materials since 2009.[198] Although online platforms typically encourage the sharing of uploaded content, the notice-and-takedown procedures in their terms of use may simultaneously discourage sharing as a means of reducing their own legal risk associated with copyright-infringement liability.[199] In other words, although Type II and Type III sharing are widely permitted over those platforms, Type I sharing is obviously not encouraged there.

## 5.2 The changing business models of platforms

### 5.2.1. Video sharing platforms

Typically, online platforms need copyright owners' licensing of certain exclusive rights, such as rights of reproduction, distribution, public display, and making content available online. In China, the most important economic right necessary for platforms to obtain licensing is the network communication right, which is a comprehensive right conceptually covering other economic rights, such as reproduction rights, in the Internet environment.[200] If these platforms do not plan to further produce

---

[196] Fenxia Yang and Huirong He, '弹幕视频网站哔哩哔哩网侵权问题辨析 (Analysis of Danmaku Video Website Bilibili's Infringement Issues)' (*People,* 2016) < http://media.people.com.cn/n1/2016/1228/c409156-28984083.html> accessed 21 January 2019;Shanghai Business Newspaper, '弹幕网站 bilibili 侵权被判赔 10 万元(Danmaku Website Bilibili Held to Be Liable for Infringement and RMB$100,000 Compensation)' (*Sohu*) < https://m.sohu.com/n/406122038/> accessed 21 January 2019.

[197] Xiao Ma, 'Establishing An Indirect Liability System for Digital Copyright Infringement in China: experience from the United States' Approach' (2015) 4 NYU J Intell Prop & Ent L 253, 270-72; Jia Wang, 'Toward a More Balance Safe Harbour Protection System for Internet Service Providers in China' (2015) 45 Hong Kong LJ 551, 868-873; Jie Wang, 'Development of Hosting ISP's Secondary Liability for primary Copyright Infringement in China—As Compared to the US and German Routes' (2015) 46 IIC 275, 284-85, 291-92, 298-99, 305-07.

[198] Priest (n 2) 1083.

[199] Alm (n 48) 123-24.

[200] Jyh-An Lee 'Copyright Divisibility and the Anticommons' (2016) 32 Am U Int'l L Rev 117, 136-37

their own content based on materials uploaded by copyright holders, they do not need licenses of right of preparing for derivative work or right of adaptation. However, YouTube, a dominant platform outside China, has developed much more aggressive copyright practices in recent years by requiring copyright owners to grant a worldwide royalty-free license of right of preparing for derivative work and right of adaptation to it when uploading videos to its platform.[201] YouTube has obtained a degree of control in excess of what is needed for a video-sharing platform.[202] Compared to individual or small and medium enterprise (SME) users, platforms with large market shares, such as YouTube, have much stronger bargaining power when asking copyright owners to share their content for the purpose of further creation.[203] Therefore, some commentators have voiced the criticism that

> the greatest drawback of the YouTube process is that copyright owners license YouTube only. The license does not 'pass through' to the user who generated the work and who may have created a derivative work. The user remains an infringer while the redistribution becomes licensed.[204]

In China, some popular platforms only require copyright owners to license network communication rights to them, whereas others employ an aggressive, YouTube-style strategy. For example, the notable video-sharing platform Youku only requires copyright holders to license their network communication right when uploading videos.[205] By contrast, copyright holders need to globally, permanently, irrevocably, and freely license their entire set of economic rights—which includes rights of adaptation, publication, translation, preparing derivative works, and

---

[201] YouTube, Term of Service, Sec. 6.C: "…by submitting Content to YouTube, you hereby grant YouTube a worldwide, non-exclusive, royalty-free, sublicenseable and transferable license to use, reproduce, distribute, prepare derivative works of, display, publish, adapt, make available online or electronically transmit, and perform the Content in connection with the Service and YouTube's (and its successors' and affiliates') business, including without limitation for promoting and redistributing part or all of the Service (and derivative works thereof) in any media formats and through any media channels…"

[202] Adam Shatzkes, 'The Destruction of An Empire: Will Viacom End YouTube's Reign?' (2010) 26 Touro L Rev 287, 299-300.

[203] Brian Day, 'The Super Brawl: The History and Future of the Sound Recording Performance Right' (2009) 16 Mich Telecomm & Tech L Rev 179, 203; John B. Meisel, 'Economic and Legal Issues Facing YouTube and Similar Internet Hosting Web Sites' (2009) 12(8) J Internet L 1, 10; Brett White, 'Viacom v. YouTube: A Proving Ground for DMCA Safe Harbors Against Second Liability' (2010) 24 St John's J Legal Comment 811, 812.

[204] DeviantART, 'Request for Comments on Department of Commerce Green Paper, Copyright Policy, Creativity, and Innovation in the Digital Economy, NO. 130927852-3852-01, Comments of DeviantART' 28 (Washington D.C., 13 Nov. 2013) <http://www.ntia.doc.gov/files/ntia/deviant_art_comments.pdf> accessed 21 January 2019.

[205] Youku, 'Terms of Uploading Service' Article 5 (*Youku*, Beijing, 1 January 2018) < http://csc.youku.com/feedback-web/help/index.html?spm=a2h4v.8841035.4646085.3&style=1&loreid=548> accessed 21 January 2019.

performance—when uploading content to the other two popular platforms, Tudou and iQiyi.[206] Because Tudou is a subsidiary of Alibaba[207] and iQiyi is *447 invested by Baidu, it is not surprising that these two Chinese Internet giants have ambitious plans to obtain licenses for the uploaded content.

It might be unclear whether the active strategy of platforms will benefit or harm the copyright-sharing economy in China. On one hand, sufficient incentives and a sustainable business model for platforms are crucial to their ability to play a key role in the sharing economy. On the other, as a critique of YouTube's aggressive copyright strategy clearly stated, copyright owners' sharing with major platforms does not help other users legally reuse the content.[208] Additionally, key platform operators such as Youku Tudou and iQiyi have become major copyright producers and owners. They have demonstrated a much more ambitious strategy for market entrance into the entertainment industry than has YouTube, their counterpart in the Western world. If they make use of the content uploaded by other copyright owners based on the aggressive license agreements mentioned above, the roles of platforms, copyright owners, and users will converge in an unprecedented way.

Nonetheless, one might not necessarily need to be pessimistic about the platforms' assertive copyright strategies and their impact on the copyright-sharing economy. All major video-streaming platforms in China have developed their own production lines of high-quality dramas or films.[209] These platforms also collaborate closely with international content providers or distributors, such as BBC, Disney, Fox, Warner Brothers, and Netflix.[210] Most major Chinese platforms have transformed into content companies.[211] Such development has dinstinguished the Chinese platforms from their counterpart YouTube in the Western world. Because those companies have transformed from pure online platforms to content companies, it is natural for them to claim copyright as much as possible in various transactions. Although Tudou, iQiyi, and their counterpart YouTube outside China have required permanently, irrevocably, and freely license from users, neither of these two companies have used UGC content to prepare for derivative works. The reality is with enormous investment in content production, the quality of content produced by those platforms by and large have been obviously superior to user-generated videos, though the latter's quality has improved substantially.[212] The original-production content offered by major video-streaming websites, such as iQiyi and Youku, has already exceeded 50 percent of their offering.[213] More and more users visit those Chinese video platforms primarily for such professionally-produced high-quality content, instead of UGC. The recent success of major Chinese video platforms, such as Youku, iQiyi, and Tencent Video, has proved that professionally-produced content is normally more attractive to audience than UGC in the context of online video.[214] Therefore, the aggressive assertion of UGC copyright does not necessarily mean that those platforms will actively use UGC to prepare for derivative works. After all, content creation and its commercialization, rather than commercial exploitation of UGC, have become these platforms' major business.

---

[206] Tudou, 'Service Agreement for Subscribers' Article 5 (*Tudou*, Shanghai, 30 June 2017) < https://new.tudou.com/about/account > accessed 21 January 2019; iQiyi, 'iQiyi PPS Internet Service Agreement for Users' Article 4.2 (*iQiyi*, Beijing) < http://www.iqiyi.com/user/register/protocol.html> accessed 21 January 2019.

[207] It should be noted that although Tudou and Youku are two different platforms, they were merged into one company Youku Tudou in 2012. The company was acquired by Alibaba in 2016 for US$4 billion. Patrick Frater, 'Alibaba Completes $4 Billion Takeover of Youku Tudou' (*Variety*, 2016) < http://variety.com/2016/biz/asia/alibaba-completes-youku-tudou-takeover-1201746908/> accessed 21 January 2019; Kazunori Takada 'Youku to Buy Tudou, Creating China Online Video Giant' (*Reuters*, 2012) < https://www.reuters.com/article/us-youku-tudou/youku-to-buy-tudou-creating-china-online-video-giant-idUSBRE82B0HD20120312> accessed 21 January 2019.

[208] Deviant ART (n204) 28.

[209] Liz Shackleton, 'Netflix Bulks Up on Chinese Content from iQiyi, Youku' (*Screen Daily*, 2017) <https://www.screendaily.com/news/netflix-bulks-up-on-chinese-content-from-iqiyi-youku/5124708.article> accessed 21 January 2019; Kaihao Wang, 'Dramatic Improvement" (*China Daily*, 2018) <http://usa.chinadaily.com.cn/a/201804/18/WS5ad68c6aa3105cdcf6518e5c.html> accessed 21 January 2019.

[210] Benjamin Cher, 'BBC and Youku Partner to Make BBC Earth Films Available in China' (*The Drum*, 2017) <http://www.thedrum.com/news/2017/06/07/bbc-and-youku-partner-make-bbc-earth-films-available-china> accessed 21 January 2019; Donna Fuscaldo, ' Alibabs's Youku Will Be Home to the Largest Disney Cartoon Collection in China' (*Investopia*, 2018) <https://www.investopedia.com/news/alibaba-bring-disney-cartoons-its-streaming-service/> accessed 21 January 2019; Shackleton (n 243).

Nevertheless, similar development can be found at the non-UGC video platform Netflix, which has increased its investment significantly to produce its own content. Starting from 2018, 85% of Netflix's spending has gone toward the production of its original shows and movies.[215] Video-streaming distribution channels in China and the western world have likewise expanded to the copyright content production industry. The difference is that Netflix has nothing to do with the UGC whereas most of the video-streaming website started their business from UGC content and still maintain space for users to upload their self-produced content. Moreover, while Chinese video platforms still primarily focus on the Chinese-speaking market, Netflix has targeted at multiple countries with different languages.

The shift of business model is a result of market competition between major platforms. Those platforms have realized that quality and exclusive content, rather than purely content sharing media, is the key to keep viewers,[216] not to mention that content shared by users may sometimes lead to copyright infringing liabilities. Nowadays inferior-quality (e.g. videos with low resolution or those frequently interrupted by commercials) or incomplete content can typically be viewed for free over those platforms. Major platforms, such as iQiyi, normally share the first episode for free as a sample product to attract for paid subscription.[217] Nevertheless, if viewers want to enjoy advertisement-free videos with higher quality or integrity, they are required to pay for such content. To date, major platforms such as Youku and iQiyi have had increasing revenue from pay-per-view and subscription orders.[218] Such development suggests that platforms may use UGC and inferior or incomplete content as means to attract viewers to pay for the premium content, which is a classic example of Type III sharing. When uploading videos to the platforms or browsing user-generated videos on those platforms, users may unintentionally be attracted to the professionally high-quality content **\*448** produced or acquired by the platforms. In this sense, major Chinese platforms have successfully attracted Internet traffic and potential paid consumers by providing a medium for content sharing. In other words, Type II sharing can also be used in the Type III sharing as part of the freemium model.

---

[211] Priest (n 2) 1085.
[212] Priest (n 15) 522.
[213] Montgomery & Priest (n 15) 351; Lizi, '2018 年 "优爱腾" 剧集大战 (Episode War between Youku, iQiyi and Tencent 2018)' (Diankeji, 23 Vov 2018) <http://www.diankeji.com/yule/40060.html>.
[214] Montgomery & Priest (n 15) 347.
[215] John Lynch, 'Netflix Will Have Over 1,000 Original TV Shows and Movies on its Service by the End of the Year' (2018) <https://www.businessinsider.com/netflix-over-1000-original-tv-shows-movies-by-end-of-year-2018-5> accessed 21 January 2019.
[216] Priest (n 2) 1083-84.

Major Chinese video platforms have successfully shifted from Type I sharing model to Type III model. The convergence of video-streaming platforms and copyright holders and its consequence are yet to be explored. On the one hand, based on the increasing licensing fees collected from video works since 2009, commentators have argued that copyright has been better protected by video platforms for their own interests.[219] Therefore, the convergence has created a revenue windfall for other domestic and international copyright owners, who now benefit from the platforms' subscription or pay-per-view models. This may eventually lead to more investment for professionally-produced high-quality content to be distributed in country although unlicensed and infringing content remains abundant on smaller websites. On the other hand, those video-streaming platforms certainly have increasing market power in markets of both video copyright and video distribution. The merger of two major video platforms Youku and Toudu in 2012 precisely signals such emerging market power that did not exist in the past. This new business landscape may bring about some new competition law issues, especially given the fact that major video-streaming platforms are invested by the famous Internet giants BAT, namely Baidu, Alibab, and Tencent, offering a wide range of internet services in the country, such as search engine, electronic commerce platforms, online payment, social media, and so on. Any of these companies may leverage their market power of various Internet services into the market of video-streaming, which might eventually reinforce the above mentioned market power resulting from the convergence of copyright owners and online platforms. With such market power and significant resources, Alibaba , Youku, and Tencent have actively invested in film projects for theatrical release as well. As a result, those video platforms have blurred the lines between Internet and traditional film making.

---

[217] Montgomery & Priest (n 15) 351.
[218] Ibid 1085-85.
[219] Montgomery & Priest (n 15) 348.

## 5.2.2. Text-sharing platforms

Tudou and iQiyi are not the only platforms aggressively seeking copyright from users. In the context of literary works, Baidu Baike (https://baike.baidu.com/) is a Chinese-language, collaborative, online encyclopedia established and operated by Baidu. As with Wikipedia, users volunteer to write and amend the entries and share them on Baidu Baike. However, the copyright policies of Baidu Baike are significantly different from those of Wikipedia. Contributors of Wikipedia entries are required to license their work to the public via the CC Attribution-ShareAlike 3.0 license.[220] Anyone can therefore use Wikipedia entries freely if he or she complies with the license by providing appropriate attribution and licensing derivative works under a compatible open license,[221] which is an ideal model of Type II sharing. By contrast, all the entries on Baidu Baike appear with the copyright symbol '©2018 Baidu', indicating that Baidu owns the copyright of all its user-generated entries. A possible explanation of this difference is that Wikipedia is hosted by the Wikimedia Foundation, a nonprofit organisation,[222] whereas Baidu Baike operates in a corporate setting. Therefore, Baidu Baike is more eager to claim exclusive rights to user-generated entries for potential financial gains on the platform. This is another example of Chinese online platforms' aggressive copyright approach to UGC. Although Baidu has not yet monetized online encyclopedia, the company can legally do so by publishing the user-generated entries in different mediums.

Platforms also play an important role in online fictions. The unauthorized sharing of online novels has been quite common even on some well-known platforms. For example, the Post Bar, sponsored by Baidu, has been notorious for pirating literary works from other pay-to-read websites.[223] Readers can read most of the pay-to-read content on Post Bar for free.[224] Post Bar even hires a team to type pirated version of fictions and share them online.[225] Unfortunately, copyright enforcement against such infringement has been quite weak.[226] The case of Post Bar also illustrates the Internet giant Baidu's differentiated copyright strategies toward different categories of copyright works. While Baidu has adopted Type II sharing model in the online encyclopedia market, the company has profited from Type I sharing of infringing materials in the market for online fictions. As mentioned in Section 5.2.1 of this paper, Baidu's online video platform iQiyi has transformed into a copyright-compliant and Type III sharing business model in which infringing materials are rarely seen, which is different from the case of Post Bar. A takeaway lesson here is that a firm might have diverse copyright strategies toward different copyright markets. The different strategies are resulted from competition in the industry, nature of the products, and other profit concerns.

By contrast, popular fiction platforms, such as Qidian Chinese Net (https://www.qidian.com/), have adopted the Type III freemium sharing model similar to that of video platforms analysed previously. Readers have free access to all stories on the first twenty days of subscription or to the first instalment of stories at Qidian Chinese Net.[227] If they decide to continue reading, they need to pay between RMB$ 0.02 and 0.07 to read each new installment.[228] Nevertheless, different from the video-sharing platforms, text-sharing platforms in China do not produce the content by themselves. They still rely on individual authors to produce online literature. This is because **\*449** the creation of fictions is quite different from that of videos. The best novel can be written by individual talent while it is quite costly to produce high-quality episodes or films. Platforms can rely on individual authors who create the best novels, but individual creators typically do not have sufficient resources to deliver first-class videos with excellent storylines, actors, sceneries, etc.

Although platforms normally share revenue they make from readers with authors, the majority of authors make little money from those platforms.[229] Most authors view platforms as tool to build reputation, with which they can possibly make a fortune by collaborating with other media companies to publish print editions or to adapt their work to films or computer games.[230] Compared to video-sharing platforms, fiction sharing platforms are less aggressive in claiming copyright over derivative use of the subject works. Moreover, similar to musicians and their Type III sharing strategies analysed in Section 3.3 of the paper, sharing has been a strategy for fiction authors to build reputation, with which they may further monetize their copyright work.

### 5.2.3. Music platforms

Similar to the video-sharing model mentioned above, platforms' expansion into the content-creation market has also occurred in the music industry. For example, Xiami, an Alibaba platform providing a streaming service, has started to sign independent artists who would release albums through Xiami Music.[231] This development has had a fundamental impact on the tripartite structure of copyright ecology, consisting of copyright owner (or author), copyright users, and online platforms. It has been widely recognized that digital technologies have facilitated the convergence of authors and copyright users.[232] Nonetheless, the ongoing convergence of online platforms and copyright owners in China is a quite unique phenomena from a comparative perspective. It is rare to see UGC platforms converging with or transforming into copyright holders with significant market share. The dominant UGC website outside China, YouTube, has not moved toward the business of self-produced copyright content.

---

[227] Montgomery & Priest (n 15) 345.
[228] Ibid 346.
[229] Ibid.
[230] Ibid.
[231] Tang and Lyons (n 15) 359.

Another noticeable development is the copyright management model of music platforms. Similar to the video-sharing platforms, music-streaming providers, such as Tencent, NetEase, Kugou, and Alibaba, have shifted to a model highly values exclusive licenses. Although online music infringement from Type I sharing is still widespread, those major music platforms have sought to secure exclusive licenses from major music labels, such as Universal Music, Sony, and Warner,[233] especially after the NCAC released the "Announcement Regarding Mandating Online Music Service Providers to Stop Disseminating Unlicensed Music Work."[234] In recent years, those companies have viewed the exclusively licensed works as their main competitive assets and started to sue one another based on exclusively licensed copyright.[235]

Music platforms' pricing models are normally designed in corresponding to the marketing strategy of the copyright owners as well as the platforms themselves. For example, platforms normally provide users with limited volume of free-trial music which was published not quite recently.[236] However, users need to pay for more popular music or music with better quality. This is another example of the Type III sharing. Copyright holders can typically obtain 40% or more of the platforms' revenue collecting from the subscription fees.[237] Therefore, some commenters argue that those music platforms have virtually become music copyright management companies.[238] Compared to the video-streaming platforms, most music platforms in China are less interested in developing their own copyright content and in providing space for UGC content. They focus more on developing various value-added services for their

---

subscribers. For example, QQ Music provides the best sound quality for the audio systems embedded in 11 different commodity cars, such as Mercedes, BMW, and Ford.[239] Its users can also enjoy cloud storage for around 1200 songs and priority to buy their preferred concert tickets.[240] Additionally, Kugou and QQ Music have both enabled its users to have online duets or chorus with their friends.[241] These are all examples of creative music service provided by Chinese music platforms.

**5.3 Governmental Sharing Platforms**

In recent years, Chinese government agencies have built some platforms on which to share information for specific goals. These platforms are designed either to share others' work or the government's own information. To promote open access to **\*450** scientific knowledge, the Chinese government has built several platforms on which scientists can share their research. For example, Science Paper Online (SPO) (http://www.paper.edu.cn/) is supported by the Ministry of Education and aims to provide an efficient channel for scientists to exchange their research outcomes.[242] The Chinese Academy of Sciences Institutional Repository Grid (CASIR) (http://www.irgrid.ac.cn/) is an integrated platform connecting 89 institutional repositories.[243] These platforms have been created partly because China's Ministry of Science and Technology and its National Science Foundation have required open-access deposit of the research outcomes from funded projects.[244]

Because all the scientific papers shared on these two platforms are copyright works, the platforms' copyright policies will affect the degree to which the works are shared. However, the copyright policies on both platforms may fail to help achieve underlying policy goals of open access. First, the CASIR website contains a Copyright Policies section, but the copyright policies have not been released and this section has been left empty.[245] Second, the SPO website mentions copyright in two different sections. In the Frequently Asked Questions (FAQ) section, the platform makes it clear that by submitting a paper to the website, copyright owners agree to nonexclusively license their copyright to the SPO website.[246] The Terms and Conditions section, on the other hand, stipulates that 'copyright of all works on the website are owned by the

---

[239] QQ Music, '什么是 QQ 音乐车载互联? (What is QQ Music Automobile Interconnection?)' <http://y.qq.com/y/static/down/car_introduce.html> accessed 21 January 2019.

[240] QQ Music, '票务 (Concert Tickets)' <https://y.qq.com/portal/piao_wu.html> accessed 21 January 2019.

[241] Kugou KTV <http://ktv.kugou.com/ktv/index.html> accessed 21 January 2019; QQ Muisc, '全民 K 歌 (Everyone Singing)' accessed 21 January 2019.

[242] SPO, 'Introduction to SPO' <http://www.paper.edu.cn/templates/introduction.shtml> accessed 21 January 2019.

[243] Xiang Ren and Lucy Montgomery, 'Open Access and Soft Power: Chinese Voices in International Scholarship" (2015) 37(3) Media, Culture & Society 394, 399.

[244] Chunli Bai, 'Promoting Research and Innovation with Open Access' (2014) 323 National Science Review 1, 1.

[245] CASAR, 'Copyright Policies' <http://www.irgrid.ac.cn/view-copyright> accessed 21 January 2019.

[246] SPO, 'FAQ' <http://www.paper.edu.cn/templates/problem.shtml> accessed 21 January 2019.

Science and Technology Development Center, Ministry of Education, and the information provider. No one is allowed to copy, link, quote, disseminate, summarize, adapt, translate, or imitate the website content without the written agreement of the Science and Technology Development Center'.[247] Obviously, the copyright ownership of papers uploaded to the SPO website is quite unclear. The FAQ section indicates that the author is still the copyright owner who licenses copyright to the platform, whereas the Terms and Conditions section states that the Science and Technology Development Center of the Ministry of Education is one of the copyright co-owners and further use of the works by others is determined completely by the Center. Moreover, the scope of prohibited behaviour, such as linking and summarising, is arguably too wide, and such restrictions may eventually hinder the open-access policy goal of disseminating knowledge. One might attempt to justify these restrictions by arguing that what has been shared on the SPO platforms is not copyright but access to copyright work. After all, it is the scientific ideas rather than their expression that the government aims to disseminate, and copyright will protect only expression, not ideas.[248] Nevertheless, this argument would be inconsistent with the development of the open-access movement, which promotes free distribution of scholarly material online.[249] After all, the idea of open access emphasises that users can legally copy, use, or distribute scholarly works without copyright owners' permission.[250]

Another type of government platform is that established for sharing government data. Open government data (OGD) has become a popular governmental practice and international movement in recent years,[251] and China is no exception. OGD has been viewed as a crucial strategy for building a 'data-driven economy' based on valuable government data.[252] China's Premier Li Keqiang clearly stated in March 2015 that government data should be public whenever possible, 'unless it is relevant to national security and privacy'.[253] The State Council announced on 13 August 2015 that it would foster OGD sharing in its 'Action Guidelines for Promoting Big Data Development'.[254] On 3 May 2017, the State Council released the 'Implementing Plan for Government Information Integrated System and Sharing', which aims to build online platforms on which the government sector can make available its machine-readable and reusable data

---

to the public.[255] Against this background, more than 20 local governments in China have built their own platforms on which to share their data.[256] However, the data-licensing policies vary significantly among these governments. As with licensing practice in the Western world,[257] the OGD platforms of Shanghai, and Guangzhou provide the highest level for sharing, which allows users to freely use, distribute, and share the data resources obtained from the platforms.[258] By contrast, some other **\*451** platforms do not provide any licensing terms over the OGD,[259] and still others have been quite restrictive and disallow users to reproduce the data.[260] Empirical research suggests that OGD licenses in mainland China tend to be more restrictive than those in Hong Kong and Taiwan.[261]

For platforms without any licensing arrangement, this might be the case because government agencies are not aware of the IP issues involved or because the data, dataset, or database is not original enough to be protected by copyright.[262] Nevertheless, this might not be the case for some OGD platforms in China. For instance, Zhejiang Provincial Government has made it clear in its OGD portal that the government owns copyright in the text, picture, audio, software and other forms of data from its portal.[263] Although the portal requires users to obtain a licence from the government for the use of data,[264] there is no standardised licence agreement on the website. As a result, users do not have legally efficient way to use OGD.

Moreover, platforms with too restrictive licensing terms may fail to achieve the OGD policy goal of unlocking the value of data by sharing. For example, the City of Beijing's government data portal does not allow users to transfer the data to any third party.[265] However, many new applications associated with OGD involve the transfer of government data. Numerous applications have been developed elsewhere to enable end users to make better use of the OGD associated with weather, transportation, etc. The Beijing government data portal further stipulates that the applications developed by users are subject to government approval, following that the government has the power to delete or block the applications.[266] This restriction, which is seldom seen in other jurisdictions, reflects the political reality of the country, which tends toward stringent government control of information and market activities. All these restrictions, however, may discourage new business models and value-added service built upon OGD.

---

[255] State Council of PRC, 'Implementing Plan for Government Information Integrated System and Sharing (政务信息系统整合共享实施方案)' (State Council, 2017)
<http://www.gov.cn/zhengce/content/2017-05/18/content_5194971.htm> accessed 21 January 2019.
[256] 'Report of Local Government Open Data Platforms in 2017 Released in Beijing (2017 中国地方政府数据开放平台报告在京发布)' (*Tencent Cloud Community*, 2017)
<https://cloud.tencent.com/developer/article/1043792> accessed 21 January 2019.
[257] Lee (n 251) 217-26.
[258] Shanghai Municipal Government Data Service Net, 'Term of Use" (2015) <
http://www.datashanghai.gov.cn/home!toUseProvisions.action> accessed 28 April 2018; Guangzhou Municipal Integrated Open Government Data Platform, 'Disclaimer' <
http://datagz.gov.cn/odweb/sitelaw.htm> accessed 21 January 2019.
[259] For example, the Chongqing OGD platform (http://www.cqdata.gov.cn/) does not provide any licensing arrangement for the released data.
[260] For example, the Zhejiang OGD platform does not permit reproduction of the released data. Zhejiang Government Service Network, 'Website Statement'
<http://www.zjzwfw.gov.cn/col/col42277/index.html> accessed 21 January 2019.
[261] Jyh-An Lee, 'Open Government Data Licences in the Greater China Region' in Susan and Jessica C Lai (eds), *Making Copyright Work for the Asian Pacific: Juxtaposing Harmonisation with Flexibility* (ANU Press 2018) 215.
[262] Lee (n 251) 229-30; Li and Greenleaf, (n 19) 151-52.
[263] Zhejiang Provincial Government, 'Statement of the Website (网站声明)'
<http://www.zjzwfw.gov.cn/col/col42277/index.html> accessed 21 January 2019.
[264] Ibid.

The Chinese government's information-sharing platforms have some important implications. First, the government builds platforms only for some categories of content with certain policy goals, such as open access and open data. Most online-sharing platforms for copyright or copyright work are still initiated by the private sector. Second, one of the government's major functions is to enhance social welfare by supplying public goods that the market fails to produce.[267] The governmental sharing platforms described above can be viewed as a form of public-goods provision. The private sector typically does not have sufficient incentives to build such platforms, which have little potential to directly generate revenue. Third, if the content shared on the platform is copyrighted work, the platform's copyright policy plays a vital role in the policy implementation. As the purpose of these sharing activities is to promote public interest, they are similar to Type II sharing. However, poorly designed licensing terms would possibly impede the benefit of such sharing. Put differently, an overly restrictive copyright policy for the platform may undermine the policy objective of open access or open data.

## 6. Conclusion

Conventional wisdom suggests that copyright owners exert exclusive rights over their works and are unwilling to share their copyright without reasonable compensation. However, the sharing economy enabled by digital technologies has provided an increasing number of exceptions to the proprietary use of copyright in various settings. This paper has investigated dynamic incentives for Chinese copyright owners, users, and online platforms to participate (or not to participate) in the sharing economy. As in many other jurisdictions, the group of users in China has always been ready to embrace the sharing economy without hesitation. Nevertheless, users are typically the least resourceful group in the copyright ecology, and their behaviours are usually subject to high risk of copyright infringement associated with Type I sharing.

---

[265] City of Beijing, 'Disclaimer (免责声明)' <http://www.bjdata.gov.cn/gywm/mzsm/index.htm> accessed 21 January 2019.
[266] Ibid.
[267] Joseph E. Stiglitz, Economics of the Public Sector (3rd edn, W.W. Norton & Company, 2000) 79-80.

Therefore, the scope of users' participation in the Type I sharing depends on the range of limitations or exceptions to copyright. From a statutory viewpoint, Article 22 of the CCL provides an exhaustive list of exceptions that provides less room for users than does the U.S. Copyright Act. Recent judicial developments seem to favour the open approach, which is similar to the U.S. fair-use doctrine. Nevertheless, whether the Chinese judiciary will completely embrace the open approach and what the appropriate standard for this open approach is remain unclear.

Copyright owners and online platforms have more opportunities than copyright users to shape a new sharing economy for their own interests. In China, even though F/OSS communities have become a successful example of innovation through the combination of Type II and Type III sharing, the majority of copyright owners and copyright industries remain reluctant to embrace the sharing model. In the music industry, Type III sharing model has achieved preliminary success in the form of streaming services. Some artists emerged from obscurity and became recognisable pop singers by sharing their music online. Both F/OSS and the music industry have demonstrated that a sustainable sharing model must align the interests of businesses and individual users.

Online platforms are probably the most noticeable case in the copyright-sharing economy in China. Chinese online **\*452** platforms and the music industries have developed new business models capable of monetizing content partly based on Type III copyright sharing. As China's cultural and creative industries become more international and commercially focused, its online content sharing economy has transformed from an infringement-based Type I model to the copyright-orient or copyright-compliant Type III model. Consumers now can rarely see infringing content for free on the major video-streaming websites. A takeaway lesson from this Chinese experience is that only when the platform operators realize that copyright compliance is for their own benefit, will sharing take place in a legal way and copyright then be better protected. Moreover, firms, such as Baidu, might implement Types I, II, and III sharing models simultaneously for different categories of copyright works because of different industrial competition landscape, nature of the products or services, and other profit concerns.

Additionally, the Chinese government has built certain platforms on which to share information for policies such as open science and OGD. Compared to commercial platforms, these government-supported platforms create less complicated and far fewer copyright controversies. Nonetheless, those governmental sharing platforms should ensure that their copyright policies are consistent with the policy goal of open access or open data. Poorly designed copyright practices on a platform may prevent a magnificent policy goal from being realised.

**Acknowledgement**

The author is grateful to anonymous reviewers, Liyang Hou, Richard (Luqiang) Lin, Weijie Huang, Laura Pedraza-Farina, Sophie Stalla-Bourdillon, Chunyan Wang, and anonymous reviewers for their helpful comments and suggestions. The author would also like to thank Yangzi Li, Lesley Luo Jingyu and Xiaoding Zhuo for their extraordinary research assistance. This study was supported by a grant from the Research Grants Council in Hong Kong (Project No.: CUHK 14612417).